\documentstyle[preprint,prd,aps,epsfig]{revtex}
\begin{document}
\draft
%
\title{Cosmological Equations for a Thick Brane}
\author{Philippe Mounaix}
\address{Centre de Physique Th\'eorique, \\
UMR 7644 du CNRS, \\
Ecole Polytechnique, 91128 Palaiseau Cedex, France.}
\author{David Langlois}
\address{Institut d'Astrophysique de Paris, \\
UPR 341 du CNRS, \\
98bis Boulevard Arago, 75014 Paris, France}
\date{\today}
\maketitle
\begin{abstract}
Generalized Friedmann equations governing the cosmological evolution
inside a thick brane embedded in a five-dimensional Anti-de Sitter
spacetime are derived. These equations are written in terms of
four-dimensional effective brane quantities obtained by integrating,
along  the fifth dimension, over the brane thickness. In the case of a
Randall-Sundrum type cosmology, different limits of these effective
quantities are considered yielding cosmological equations which
interpolate between the thin brane limit (governed by unconventional
brane cosmology), and the opposite limit of an ``infinite''  brane
thickness corresponding to the familiar Kaluza-Klein approach. In the
more restrictive  case of a Minkowski bulk, it is shown that no
effective four-dimensional reduction is possible in the regimes where
the brane thickness is not small enough.
\end{abstract}
\pacs{PACS numbers:}
\section{Introduction}\label{sec1}
The idea of extra-dimensions where ordinary matter is confined within a
lower dimensional submanifold has received an enormous amount of
attention during the last few years (see e.g. \cite{rubakov01} for a
recent review). This idea has been explored, in particular, in the
cosmological context.
\paragraph*{}In the case of a single extra-dimension, a simplifying
assumption has been to consider our brane-universe as infinitely thin
along this extra-dimension. It has been shown that this approximation
is quite valid as long as the energy scales one wishes to explore are
much smaller that the energy scale related to the inverse thickness of
the brane.
\paragraph*{}The purpose of the present paper is to explore the
modification of the equations yielding  the cosmological evolution
within the brane when the finite thickness of the brane is taken into
account. We consider this question in a purely phenomenological
approach by assuming that the matter energy-momentum tensor describing
the brane has some distribution over a finite length interval along the
fifth dimension. Our approach (similar to that of \cite{kkop99}) is
thus more crude, but also more general, than the domain wall
description of a brane, for which the brane is embodied by a scalar
field, such as in \cite{cg99,dfgk00} (see also \cite{bcg99} for a
recent study in 4d gravity and \cite{cs97} for a review on domain
walls, in particular in supergravity).
\paragraph*{}Two limiting cases are already known. The case of an
infinitely thin brane gives the unconventional Friedmann equation of
\cite{bdl99}, which can be made compatible with ordinary cosmology at
late times if one introduces both a negative cosmological constant in
the bulk spacetime and a related tension (unmeasurable by cosmological
observers) in the brane \cite{cosmors,sms99,bdel99,ftw99}. The other and
opposite limit is that of an infinitely thick brane, in a sense defined
below. This limit effectively corresponds to a Kaluza-Klein picture
where matter is homogeneously distributed over the extra-dimension(s)
and, as it has been known for a long time, ordinary Friedmann equations
are recovered. Our analysis gives access to a description of the
intermediate cases.
\paragraph*{}The paper is organized as follows. Our effective Friedmann
equation is established in Section\ \ref{sec2} and written in terms of
four-dimensional effective quantities. Section\ \ref{sec3} is devoted
to the cosmology for a  brane with a tension,  the cosmological  effect
of which is cancelled by a negative bulk cosmological constant. We
consider the limits of a low brane energy (for all brane thickness) and
of a small brane thickness (for all brane energy), and we propose
uniform expressions for the cosmological equations valid over the whole
parameter domain covered by these two limits. In Section\ \ref{sec4},
we determine the higher order corrections (in the normalized brane
thickness) to the effective cosmological equations in the simpler case
of a Minkowski bulk. In particular, we show that no effective
four-dimensional reduction is possible in the regimes of order one
normalized brane thickness.
%
%
\section{Effective Friedmann equation for a thick brane}\label{sec2}
We consider a five-dimensional spacetime, which is homogeneous and
isotropic along three spatial dimensions, and which contains a thick
brane.  For simplicity, we will restrict ourselves to the case of a
mirror symmetric brane (and thus spacetime) along the extra-dimension.
It is then  always possible to find a Gaussian coordinate system,
starting from the hypersurface representing the ``center'' of the
brane, in which the line element takes the simple diagonal form
\begin{equation}\label{eq2.1}
ds^2\equiv g_{AB}dx^Adx^B
=-n^2(t,y)dt^2+a^2(t,y)\delta_{ij}dx^idx^j+r_b^2dy^2,
\end{equation}
where $y$ is the coordinate of the fifth dimension and $r_b$ is the
brane thickness which will be assumed to be time independent. The brane
is localized  between $y=-1/2$ and $y=1/2$. To be compatible with the
spacetime symmetries, the energy-momentum tensor of the matter content
in the brane is necessarily of the form
\begin{equation}\label{eq2.2}
{T^A}_B={\rm diag}\left\lbrack\frac{1}{r_b}
\left(-\rho_b,p_b,p_b,p_b\right),P_T\right\rbrack,
\end{equation}
where $\rho_b$, $p_b$, and $P_T$ are functions of $t$ and $y$. The
presence of a negative cosmological constant in the bulk, $\Lambda$, is
accounted for by an energy-momentum tensor of the form
\begin{equation}\label{eq2.3}
{T^A}_B\vert_{{\rm Bulk}}=-\Lambda {g^A}_B.
\end{equation}
Note that, strictly speaking, $\Lambda$ as defined here does not have
the dimension of a cosmological constant but rather that of a
five-dimensional energy density.
\paragraph*{}The purpose of this paper is to establish effective
cosmological equations for an observer living in the brane. Due to the
finite thickness of the brane, there is some arbitrariness in the
definition of what the effective four-dimensional quantities should be.
The simplest prescription one can think of consists in defining the
four-dimensional effective quantity associated to a five-dimensional
quantity $Q(t,y)$ as its spatial average over the brane thickness
\begin{equation}\label{eq2.4}
\langle Q\rangle(t) =\int_{-1/2}^{1/2}Q(t,y)\, dy.
\end{equation}
In this paper, we will adopt this simple prescription.
The four-dimensional ``observable'' counterparts of $a$, $\rho_b$, and
$p_b$ are thus $\langle a\rangle =\int_{-1/2}^{1/2}a\, dy$,
$\langle\rho_b\rangle =\int_{-1/2}^{1/2}\rho_b\, dy$, and $\langle
p_b\rangle =\int_{-1/2}^{1/2}p_b\, dy$. Note that, in contrast with the
usual dimensional reduction in field theory, we do not integrate over
the whole (compact) extra-dimension but only over the extension of the
brane in the extra-dimension.
\paragraph*{}The next step is now to obtain evolution equations for the
four-dimensional effective quantities from the five-dimensional
Einstein's equations
\begin{equation}\label{eq2.5}
G_{AB}\equiv R_{AB}-{1\over 2}R\ g_{AB}=\kappa_{(5)}^2 T_{AB},
\end{equation}
where $R_{AB}$ is the Ricci tensor and $R$ its trace. The components of
the Einstein equations, with the metric (\ref{eq2.1}) and the
energy-momentum tensors (\ref{eq2.2}) and (\ref{eq2.3}) can be obtained
from the components of the Einstein tensor given in the appendix. In
particular, the 05 component of the Einstein equations yields
\begin{equation}\label{eq2.6}
n(t,y)=\xi(t)\dot{a}(t,y),
\end{equation}
where the dot stands for a partial derivative with respect to time $t$
and $\xi$ depends on the normalization prescription for $n$. In the
following, we will take the normalization $\langle n\rangle =1$ which
gives $\xi =\dot{\langle a\rangle}^{-1}$. The 00 component of the
Einstein equations then reads
\begin{equation}\label{eq2.7}
(a^2)''=-\frac{2\kappa_{(5)}^2}{3}
\left(r_b\rho_b +r_b^2\Lambda\right) a^2
+2r_b^2\dot{\langle a\rangle}^2,
\end{equation}
where a prime denotes a partial derivative with respect to the
coordinate $y$. Integrating this equation over the brane, one obtains
\begin{eqnarray}\label{eq2.8}
a(1/2)'&=&\frac{1}{a(1/2)}\left\lbrack
-\frac{\kappa_{(5)}^2}{6}r_b\langle\rho_b a^2\rangle
-\frac{\kappa_{(5)}^2}{6}r_b^2\Lambda
\langle a^2\rangle
+\frac{1}{2}r_b^2\dot{\langle a\rangle}^2\right\rbrack
\nonumber \\
&=&\frac{\langle a\rangle}{\alpha}\left\lbrack
-\varepsilon\eta
-\frac{r_b^2}{\ell_{\Lambda}^2}\tilde{\eta}
+\frac{1}{2}r_b^2H_{\langle a\rangle}^2
\right\rbrack ,
\end{eqnarray}
where we have introduced the effective four-dimensional Hubble parameter
\begin{equation}\label{eq2.9}
H_{\langle a\rangle}
\equiv\dot{\langle a\rangle}/\langle a\rangle ,
\end{equation}
the Anti-de Sitter lengthscale associated to the (negative) cosmological
constant in the bulk
\begin{equation}\label{eq2.10}
\ell_{\Lambda}\equiv\sqrt{-\frac{6}{\kappa_{(5)}^2\Lambda}},
\end{equation}
and the dimensionless quantities
$\varepsilon$, $\alpha$, $\eta$, and $\tilde{\eta}$, respectively
defined by
\begin{equation}\label{eq2.11}
\varepsilon\equiv\frac{\kappa_{(5)}^2}{6}r_b \langle\rho_b\rangle,
\end{equation}
and
\begin{equation}\label{eq2.12}
\alpha\equiv\frac{a(1/2)}{\langle a\rangle},\ \ \ \ \ \
\eta\equiv\frac{\langle\rho_b a^2\rangle}{\langle\rho_b\rangle\langle
a\rangle^2},\ \ \ \ \ \
\tilde{\eta}\equiv\frac{\langle a^2\rangle}{\langle a\rangle^2}.
\end{equation}
Whereas $\varepsilon$ characterizes the thickness of the brane, the quantities
$\alpha$, $\eta$, and $\tilde{\eta}$ characterize the inhomogeneity of
the brane along the fifth dimension (in the case of a homogeneous
brane, one has $\alpha=\eta=\tilde{\eta}=1$). The 55 component of the
Einstein equations can be written as
\begin{equation}\label{eq2.13}
\dot{F}=\frac{2}{3}\kappa_{(5)}^2\dot{a}a^3P_T,
\end{equation}
with $F=a^2\lbrack ({a'}/r_b)^2-\dot{\langle a\rangle}^2
+\kappa_{(5)}^2a^2\Lambda /6\rbrack$. Imposing the boundary condition
$P_T(\pm 1/2)=0$, one has $\dot{F}(\pm 1/2)=0$ which gives, after
time-integration,
\begin{equation}\label{eq2.14}
r_b^2H_{\langle a\rangle}^2=
\left\lbrack\frac{a'(1/2)}{\langle a\rangle}\right\rbrack^2
+\frac{{\cal C}r_b^2}{\alpha^2\langle a\rangle^4}
+\frac{\kappa_{(5)}^2}{6}\alpha^2r_b^2\Lambda ,
\end{equation}
where ${\cal C}$ is a constant of integration.
Injecting Eq.\ (\ref{eq2.8}) into Eq.\ (\ref{eq2.14}), one obtains
\begin{equation}\label{eq2.15}
\alpha^2r_b^2H_{\langle a\rangle}^2=
\left\lbrack -\varepsilon\eta
+{r_b^2\over \ell_{\Lambda}^2}\tilde{\eta}
+\frac{1}{2}r_b^2H_{\langle a\rangle}^2\right\rbrack^2
+\frac{{\cal C}r_b^2}{\langle a\rangle^4}
-{r_b^2\over \ell_{\Lambda}^2}\alpha^4.
\end{equation}
This equation is one of the main results of this paper. It provides the
generalization of the (first) Friedmann equation relating the effective
four-dimensional Hubble parameter to the effective brane energy density
and the brane thickness. It of course also includes some dependence on
the exact profile of the energy-momentum tensor but this dependence has
been conveniently reduced to the three dimensionless quantities\
(\ref{eq2.12}), which in typical cases are of order unity. From
Eqs.\ (\ref{eq2.7}) and\ (\ref{eq2.15}), one finds that a necessary
condition for the brane to be homogeneous along the fifth dimension
reads
\begin{equation}\label{eq2.16}
\frac{{\cal C}}{\langle a\rangle^4}=
\frac{2\varepsilon}{r_b^2}
-\frac{1}{\ell_{\Lambda}^2},
\end{equation}
which, in general, cannot be fulfilled uniformly in time unless the
energy-momentum content of the brane reduces to a mere brane tension
(as e.g. in the original static Randall-Sundrum model \cite{rs99b}). It follows
that, in the cosmological context, a brane with a finite thickness
cannot remain homogeneous along the fifth dimension.
\paragraph*{}Equation (\ref{eq2.15}) can be rewritten in the form
\begin{eqnarray}\label{eq2.17}
&&H_{\langle a\rangle}^2=
\frac{2}{r_b^2}
\left(\alpha^2
+\varepsilon\eta
-\frac{r_b^2}{\ell_{\Lambda}^2}\tilde{\eta}\right)
\nonumber \\
&&\times
\left\lbrack
1\pm\sqrt{1-\frac{(
\varepsilon\eta
-\tilde{\eta}r_b^2/\ell_{\Lambda}^2)^2
+{\cal C}r_b^2/\langle a\rangle^4
-\alpha^4r_b^2/\ell_{\Lambda}^2}
{(\alpha^2
+\varepsilon\eta
-\tilde{\eta}r_b^2/\ell_{\Lambda}^2)^2
}}\right\rbrack ,
\end{eqnarray}
which admits real solutions if the arbitrary constant ${\cal C}$ fulfills
the condition
\begin{equation}\label{eq2.18}
\frac{{\cal C}r_b^2}{\langle a\rangle^4}\le
2\alpha^2\left(\varepsilon\eta
-\frac{r_b^2}{\ell_{\Lambda}^2}\tilde{\eta}\right)
+\alpha^4\left( 1+\frac{r_b^2}{\ell_{\Lambda}^2}\right) .
\end{equation}
In the following, we will consider Expression\ (\ref{eq2.17}) with
the minus sign only, which corresponds to a well defined cosmology in
the $r_b\rightarrow 0$ limit. It is important to notice that this
equation should be regarded as an implicit equation for $H_{\langle
a\rangle}$ because $\alpha$, $\eta$, and $\tilde{\eta}$ can depend on
$H_{\langle a\rangle}$.
\paragraph*{}To conclude this section, it can be instructive to consider
the two opposite limits of small and large $r_b$. In the limit
$r_b\rightarrow 0$, Equation\ (\ref{eq2.17}) reduces to
\begin{eqnarray*}
H_{\langle a\rangle}^2&\simeq &\frac{1}{\alpha^2}
\left(\frac{\varepsilon^2\eta^2}{r_b^2}
+\frac{{\cal C}}{\langle a\rangle^4}
-\frac{\alpha^4}{\ell_{\Lambda}^2}\right) \\
&=&\frac{1}{\alpha^2}
\left(\frac{\kappa_{(5)}^2}{36}\langle\rho_b\rangle^2\eta^2
+\frac{{\cal C}}{\langle a\rangle^4}
+\frac{\kappa_{(5)}^2}{6}\Lambda\alpha^4\right).
\end{eqnarray*}
It will be shown in the following that, in this limit, both $\alpha$ and
$\eta$ tend to $1$. Thus, one exactly recovers the unconventional
Friedmann equation of thin brane cosmology, where the brane energy density
enters quadratically \cite{bdl99,bdel99}. In the opposite limit
$r_b\rightarrow +\infty$, Equation\ (\ref{eq2.17}) reduces to
$$
H_{\langle a\rangle}^2\simeq
\frac{2\varepsilon\eta}{r_b^2}-\frac{2\tilde{\eta}}{\ell_{\Lambda}^2}+
\frac{2}{r_b^2}\sqrt{\frac{\alpha^2(\alpha^2-2\tilde{\eta})}{\ell_{\Lambda}^2}
-\frac{{\cal C}}{\langle a\rangle^4}},
$$
where terms up to $O(r_b^{-1})$ have been kept. In the case of a
Minkowski bulk, $\ell_{\Lambda}^{-2}=0$, Eq.\ (\ref{eq2.18}) yields
${\cal C}/\langle a\rangle^4\rightarrow 0$ as $r_b\rightarrow +\infty$
and one obtains
$$
H_{\langle a\rangle}^2\simeq\frac{2\varepsilon\eta}{r_b^2}
=\frac{\kappa_{(5)}^2\eta}{3r_b}\langle\rho_b\rangle,
$$
which corresponds to the standard Friedmann equation, in which the
energy  density enters linearly with the effective four-dimensional
gravitational constant $\kappa_{(4)}^2\equiv\kappa_{(5)}^2\eta/r_b$.
Note that, in this limit, both sides of the necessary condition\
(\ref{eq2.16}) vanish, which allows the possibility of a homogeneous
brane along the fifth dimension ($\eta =1$), in agreement with the
usual Kaluza-Klein picture.
%
%
\section{Randall-Sundrum type cosmology of a thick brane}\label{sec3}
In thin brane cosmology, the unconventional Friedmann equation can be
made compatible with the current cosmological observations if one
assumes that the brane energy density consists of ordinary cosmological
matter living in the brane and of a tension $\lambda$ adjusted so as to
compensate the negative cosmological constant in the bulk
\cite{cosmors,bdel99}. This adjustment corresponds to the fine-tuning
condition of Randall-Sundrum \cite{rs99b}, which can be reexpressed as
the simple relation
\begin{equation}\label{eq3.1}
\ell_{\lambda}=\ell_{\Lambda},
\end{equation}
where the lengthscale $\ell_{\lambda}$ is defined from the brane tension
$\lambda$ as
\begin{equation}\label{eq3.2}
\ell_{\lambda}\equiv\frac{6}{\kappa_{(5)}^2 \lambda}.
\end{equation}
In this section, we will also assume that the energy-momentum content of
the brane can be decomposed into two parts,
\begin{mathletters}
\begin{eqnarray}
&&\rho_b(t,y)=\lambda+\rho_m(t,y), \label{eq3.3a} \\
&&p_b(t,y)=-\lambda+p_m(t,y), \label{eq3.3b}
\end{eqnarray}
\end{mathletters}
where $\lambda$, which is strictly constant, corresponds to a tension of
the brane. For the sake of simplicity, we assume here that $\lambda$ is
independent of $y$. The case of a non-trivial profile of the brane
tension along the fifth dimension could also be easily considered by
writing $\lambda = \langle\lambda\rangle +\delta\lambda (y)$ and by
defining an effective ``matter'' energy-momentum content
$\tilde{\rho}_m(t,y)=\rho_m(t,y)+\delta\lambda (y)$ and
$\tilde{p}_m(t,y)=p_m(t,y)-\delta\lambda (y)$. Since
$\langle\delta\lambda (y)\rangle =0$, it can be seen that the only
modification induced by such a brane tension profile reduces to adding
the geometrical term $\langle\delta\lambda a^2\rangle
/(\langle\lambda\rangle\langle a\rangle^2)$ to the parameter
$\varepsilon_{\rho}$ defined in Eq.\ (\ref{eq3.6}). We will not analyse
further on the effects of this extra term and restrict ourselves to the
case of a $y$-independent brane tension. Injecting Eq.\ (\ref{eq3.3a})
into the effective Friedmann equation\ (\ref{eq2.17}), it can be easily
seen that the generalization of the Randall-Sundrum fine-tuning
allowing for a finite brane thickness is given by
\begin{equation}\label{eq3.4}
\ell_{\lambda}=\frac{\tilde{\eta}(1-u)}{\alpha^2}\ell_{\Lambda},
\end{equation}
where we have introduced the dimensionless ratio
\begin{equation}\label{eq3.5}
u\equiv\frac{r_b\ell_{\lambda}}{\ell_{\Lambda}^2}=
-\frac{r_b\Lambda}{\lambda}.
\end{equation}
In the limit where $r_b$ goes to zero, it can be checked that Eq.\
(\ref{eq3.4}) reduces to the Randall-Sundrum condition
(\ref{eq3.1}), as it should be. From Eqs.\ (\ref{eq2.17}),\
(\ref{eq3.3a}), and\ (\ref{eq3.4}), and introducing the
dimensionless parameters
\begin{equation}\label{eq3.6}
\varepsilon_{\lambda}\equiv
\frac{r_b}{\ell_{\lambda}},\ \ \ \ \
\varepsilon_{\rho}\equiv
\frac{\eta\langle\rho_m\rangle}{\lambda},
\end{equation}
where $\eta\equiv\langle\rho_m a^2\rangle /(\langle\rho_m\rangle\langle
a\rangle^2)$, one finds that Eq.\ (\ref{eq2.17}) (with the minus sign)
can be rewritten as
\begin{eqnarray}\label{eq3.7}
&&H_{\langle a\rangle}^2=
\frac{2}{r_b^2}\left\lbrack\alpha^2
(1+\varepsilon_{\lambda}\ell_{\lambda}/\ell_{\Lambda})
+\varepsilon_{\lambda}\varepsilon_{\rho}
\right\rbrack\nonumber \\
&&\times
\left\lbrace
1-\sqrt{1-\frac{\varepsilon_{\lambda}^2(
2\varepsilon_{\rho}\alpha^2\ell_{\lambda}/\ell_{\Lambda}
+\varepsilon_{\rho}^2
+{\cal C}\ell_{\lambda}^2/\langle a\rangle^4)}
{\lbrack\alpha^2
(1+\varepsilon_{\lambda}\ell_{\lambda}/\ell_{\Lambda})
+\varepsilon_{\lambda}\varepsilon_{\rho}\rbrack^2}}\right\rbrace .
\end{eqnarray}
%
%
\subsection{Cosmological equations in the low energy limit}\label{sec3a}
In this subsection, we will consider the low energy regime in which the
ordinary cosmological energy (and pressure) is small with respect to
the brane tension. For the sake of simplicity, we will take ${\cal
C}=0$. Consider the low energy limit in the regime defined by
$\varepsilon_{\rho}\ll\min(1,1/\varepsilon_{\lambda})$. Then, {\it at
lowest order in} $\varepsilon_{\rho}$, Eq.\ (\ref{eq3.7}) reduces
to
\begin{equation}\label{eq3a.1}
r_b^2H_{\langle a\rangle}^2=
\frac{2\varepsilon_{\lambda}^2\varepsilon_{\rho}}
{\varepsilon_{\lambda}+\ell_{\Lambda}/\ell_{\lambda}},
\end{equation}
and the quantities $\alpha$, $\eta$, and $\tilde{\eta}$ must be
determined at zeroth order in $\varepsilon_{\rho}$. In this limit, the
00 component of the Einstein equations [see Eq.\ (\ref{eq2.7})] reduces
to
\begin{equation}\label{eq3a.2}
\left(\frac{a^2}{\langle a\rangle^2} \right)''
= -4\varepsilon_{\lambda}(1 -u)
\frac{a^2}{\langle a\rangle^2},
\end{equation}
where we have neglected the term $2r_b^2H_{\langle a\rangle}^2$ which is
first order in $\varepsilon_{\rho}$. Restricting ourselves to the case
of physical interest $(1-u)>0$ [it can be checked {\it a posteriori},
cf. Eq.\ (\ref{eq3a.8b}), that this inequality is always fulfilled], and
integrating Eq.\ (\ref{eq3a.2}) with the boundary condition
$a'(y=0)=0$, one finds
\begin{equation}\label{eq3a.3}
a=\langle a\rangle\lbrack {\cal A}\cos(ky)\rbrack^{1/2},
\end{equation}
with
\begin{equation}\label{eq3a.4}
k^2=4\varepsilon_{\lambda}(1-u),
\end{equation}
and
\begin{equation}\label{eq3a.5}
{\cal A}=\langle\cos^{1/2}(ky)\rangle^{-2}.
\end{equation}
Note that here we necessarily have  $k < \pi$, and the scale factor $a$
is always positive throughout the brane, reaching the extremity of the
brane at $y=1/2$. It is possible to determine the metric outside the
brane by matching, at $y=1/2$, the scale factor and its derivative with
respect to $y$ to the general solution for the bulk with cosmological
constant, obtained in \cite{bdel99}, which is in fact the
Schwarschild-AdS solution expressed in Gaussian normal coordinates.
\paragraph*{}These expressions can be used to determine the evolution of
$\rho_m$ at lowest order in $\varepsilon_{\rho}$. Since neither ${\cal
A}$ nor $k$ depend on $t$, one simply has
\begin{equation}\label{eq3a.6}
\dot{\rho}_m+3H_{\langle a\rangle}(\rho_m+p_m)=0,
\end{equation}
from which it follows that, at this order, $\rho_m$ factorizes as
$\rho_m(y,t)=f(y)\langle\rho_m\rangle (t)$.
\paragraph*{}The effective four-dimensional brane cosmology
in the limit $\varepsilon_{\rho}\ll\min(1,1/\varepsilon_{\lambda})$ is
thus given by
\begin{mathletters}
\begin{eqnarray}
&&H_{\langle a\rangle}^2=
\frac{\kappa_{(5)}^2}{3}\left(
\frac{\eta}{r_b +\ell_{\Lambda}}\right)
\langle\rho_m\rangle, \label{eq3a.7a} \\
&&\dot{\langle\rho_m\rangle}+
3H_{\langle a\rangle}(\langle\rho_m\rangle+\langle p_m\rangle)=0,
\label{eq3a.7b}
\end{eqnarray}
\end{mathletters}
with
$$
\eta =\frac{\langle f(y)\cos(ky)\rangle}
{\langle f(y)\rangle\langle\cos^{1/2}(ky)\rangle^2},
$$
where $k$ and ${\cal A}$ are respectively given by Eqs.\ (\ref{eq3a.4}) and\
(\ref{eq3a.5}). Using the fine-tuning condition\ (\ref{eq3.4})
with $\alpha =\lbrack{\cal A}\cos(k/2)\rbrack^{1/2}$ and $\tilde{\eta}
=(2{\cal A}/k)\sin(k/2)$, one can reexpress the constant $k$ as a
function of $\varepsilon_{\lambda}$ in the parametric form
\begin{mathletters}
\begin{eqnarray}
&&k=2\tan^{-1}\sqrt{u/(1-u)},
\label{eq3a.8a} \\
&&\varepsilon_{\lambda}=\frac{\lbrack\tan^{-1}\sqrt{u/(1-u)}\rbrack^2}
{1-u},
\label{eq3a.8b}
\end{eqnarray}
\end{mathletters}
with $0\le u<1$. Note that since, in this limit, $\eta$ does not depend
on $t$, the only effect of the finite brane thickness is a modification
of the relation between $\kappa_{(5)}$ and $\kappa_{(4)}$. Namely, one
has
\begin{equation}\label{eq3a.9}
\kappa_{(4)}^2=\frac{\eta\kappa_{(5)}^2}{r_b+\ell_{\Lambda}}.
\end{equation}
Taking into account the constraint\ (\ref{eq3.4}), one has the limits
$\kappa_{(4)}^2\simeq\kappa_{(5)}^3\sqrt{\Lambda
/6}\simeq\kappa_{(5)}^4\lambda /6$ for $r_b/\ell_{\Lambda}\ll 1$, and
$\kappa_{(4)}^2\simeq\eta\kappa_{(5)}^2/r_b$ for $r_b/\ell_{\Lambda}\gg
1$. In the simplest case $f(y)={\rm cst.}$,
$\rho_m=\langle\rho_m\rangle$, and $\lim_{r_b\rightarrow
+\infty}\eta\simeq 1.094$. In Fig. 1 we have plotted $k$ as a function
of $u$ (solid line), and $\varepsilon_{\lambda}$ as a function of $u$
(dashed-dotted line). The infinitely thin limit corresponds to
$\varepsilon_{\lambda} \rightarrow 0$, in  which case $u\rightarrow 0$
and $k\rightarrow 0$. The opposite limit $\varepsilon_{\lambda}
\rightarrow +\infty$, yields $u\rightarrow 1$ and $k\rightarrow
k_{max}=\pi$. Figure 2 shows the dimensionless coefficients, ${\cal
A}$, $\alpha$ and $\tilde{\eta}$, characterizing the brane profile, as
a function of $k/k_{max}$. In the infinitely thin brane limit
($k\rightarrow 0$), these three coefficients tend to $1$. It can be
seen that $\tilde{\eta}$ stays almost constant over the full range of
$k$, with $\tilde{\eta}(k_{max})\simeq 1.094$. As the brane thickness
increases, $\alpha$ decreases down to zero for $k=k_{max}$ and ${\cal
A}$ increases up to the value ${\cal A}(k_{max})\simeq 1.719$.
\paragraph*{}All the results of this section have been obtained in the
limit $\varepsilon_{\rho} \ll\min(1,1/\varepsilon_{\lambda})$ by
keeping terms explicitely of lowest order in $\varepsilon_{\rho}$ and
of any order in $\varepsilon_{\lambda}$. The actual small parameter
appearing in the expansion of the square root on the right-hand side of
Eq.\ (\ref{eq3.7}) is
$\varepsilon_{\lambda}^2\varepsilon_{\rho}(\alpha^2\ell_{\lambda}/
\ell_{\Lambda}) \sim \varepsilon_{\lambda}^2\varepsilon_{\rho}\min
(1,1/\varepsilon_{\lambda})$. These equations are thus correct, in the
limit $\varepsilon_{\rho} \ll\min(1,1/\varepsilon_{\lambda})$, at
lowest order in $\varepsilon_{\lambda}^2\varepsilon_{\rho}\min
(1,1/\varepsilon_{\lambda})$ and at any order in
$\varepsilon_{\lambda}$ and $\varepsilon_{\Lambda}\equiv
r_b/\ell_{\Lambda}$.
%
%
\subsection{Cosmological equations for a small brane thickness}\label{sec3b}
In the limit of a small brane thickness, $\varepsilon_{\lambda}\ll 1$,
Eq.\ (\ref{eq3.7}) with ${\cal C}=0$ reduces to
\begin{equation}\label{eq3b.1}
r_b^2H_{\langle a\rangle}^2=
\frac{\varepsilon_{\lambda}^2(
2\varepsilon_{\rho}\alpha^2\ell_{\lambda}/\ell_{\Lambda}
+\varepsilon_{\rho}^2)}{\alpha^2
(1+\varepsilon_{\lambda}\ell_{\lambda}/\ell_{\Lambda})
+\varepsilon_{\lambda}\varepsilon_{\rho}},
\end{equation}
which is correct at third order in $\varepsilon_{\lambda}$. At this
order, $\alpha$ must be expressed at first order, and $\eta$ and
$\tilde{\eta}$ at zeroth order, [note that $\eta$ and $\tilde{\eta}$
behave as $1+O(\varepsilon_{\lambda}^2)$]. To determine $a$
perturbatively in $\varepsilon_{\lambda}$, we will use the following
$y$-expansion
\begin{mathletters}\label{eq3b.2}
\begin{eqnarray}
&&a(t,y)=\langle a\rangle(t)
\sum_{n=0}^{+\infty}\overline{a}_n(t)y^{2n}
\equiv\langle a\rangle(t)\Sigma_a(t,y), \label{eq3b.2a} \\
&&\rho_m(t,y)=\langle\rho_m\rangle(t)\sum_{n=0}^{+\infty}
\overline{\rho}_n(t)y^{2n}
\equiv\langle\rho_m\rangle(t)\Sigma_{\rho}(t,y). \label{eq3b.2b}
\end{eqnarray}
\end{mathletters}
Inserting Eqs.\ (\ref{eq3b.2}) in Eq.\ (\ref{eq2.7}), using the
constraint\ (\ref{eq3.4}), and neglecting the
$O(\varepsilon_{\lambda}^2)$ terms, one obtains
\begin{equation}\label{eq3b.3}
(\Sigma_a^2)''=-4\varepsilon_{\lambda}
\left(\varepsilon_{\rho}\Sigma_{\rho}+
\ell_{\lambda}/\ell_{\Lambda}\right)\Sigma_a^2.
\end{equation}
Since one needs $a$ at first order in $\varepsilon_{\lambda}$, one can
consider the constant (in $y$) component of Eq.\ (\ref{eq3b.3}) only.
One finds $\overline{a}_1=-\varepsilon_{\lambda}\overline{a}_0
(\varepsilon_{\rho}\overline{\rho}_0+\ell_{\lambda}/\ell_{\Lambda})$
and $a=\langle a\rangle\overline{a}_0[1-\varepsilon_{\lambda}
(\varepsilon_{\rho}\overline{\rho}_0+\ell_{\lambda}/\ell_{\Lambda})y^2]$.
In this expression, $\overline{\rho}_0$ must be determined at zeroth
order in $\varepsilon_{\lambda}$, which yields $\overline{\rho}_0=1$
and $\langle\rho_m\rangle=\rho_m$ solution to the usual 4D
energy-momentum conservation equation. Thus, at first order in
$\varepsilon_{\lambda}$ one has $a=\langle a\rangle
[1-\varepsilon_{\lambda}
(\varepsilon_{\rho}+\ell_{\lambda}/\ell_{\Lambda}) (y^2-1/12)]$,
$\alpha
=1-(\varepsilon_{\lambda}/6)(\varepsilon_{\rho}+\ell_{\lambda}/\ell_{\Lambda})$
and $\eta =\tilde{\eta} =1$. Inserting these expressions in Eq.\
(\ref{eq3.4}), one obtains (at first order in
$\varepsilon_{\lambda}$)
\begin{equation}\label{eq3b.4}
\frac{\ell_{\lambda}}{\ell_{\Lambda}}=
1-\frac{\varepsilon_{\lambda}(\varepsilon_{\rho}-2)}{3},
\end{equation}
and $\alpha$ reduces to
\begin{equation}\label{eq3b.5}
\alpha =1-\frac{\varepsilon_{\lambda}(\varepsilon_{\rho}+1)}{6}.
\end{equation}
\paragraph*{}From Eqs.\ (\ref{eq3b.1}),\ (\ref{eq3b.4}), and\ (\ref{eq3b.5})
it follows that the effective four-dimensional brane cosmology in the
limit $\varepsilon_{\lambda}\ll\min(1,1/\varepsilon_{\rho})$ is given,
at third order in $\varepsilon_{\lambda}$, by
\begin{mathletters}
\begin{eqnarray}
&&H_{\langle a\rangle}^2=
\frac{\kappa_{(5)}^2}{3}\left\lbrack
\frac{(3+r_b/\ell_{\lambda})+(3-4r_b/\ell_{\lambda})\langle\rho_m\rangle
/(2\lambda)}
{2r_b(1+\langle\rho_m\rangle /\lambda)+3\ell_{\lambda}}\right\rbrack
\langle\rho_m\rangle, \label{eq3b.6a} \\
&&\dot{\langle\rho_m\rangle}+
3H_{\langle a\rangle}(\langle\rho_m\rangle+\langle p_m\rangle)=0,
\label{eq3b.6b}
\end{eqnarray}
\end{mathletters}
These equations have been obtained in the limit
$\varepsilon_{\lambda}\ll \min(1,1/\varepsilon_{\rho})$, by keeping
terms explicitely of third order in $\varepsilon_{\lambda}$. The actual
small parameter appearing in the expansion of the square root on the
right-hand side of Eq.\ (\ref{eq3.7}) is
$\varepsilon_{\lambda}^2\varepsilon_{\rho}^2\max(1,1/\varepsilon_{\rho})$.
It follows that these equations are correct, in the limit
$\varepsilon_{\lambda}\ll \min(1,1/\varepsilon_{\rho})$, at third order
in $\varepsilon_{\lambda}\varepsilon_{\rho}$ for $\varepsilon_{\rho}
\ge 1$, and up to terms of order
$\varepsilon_{\lambda}^3\varepsilon_{\rho}$ for $\varepsilon_{\rho}\ll
1$. Note that in the domain
$\min(1,1/\varepsilon_{\rho})/\varepsilon_{\rho}\lesssim
\varepsilon_{\lambda}\ll \min(1,1/\varepsilon_{\rho})$, only the third
order (in $\varepsilon_{\lambda}$) correction corresponding to the term
$\varepsilon_{\lambda}\varepsilon_{\rho}$ must be kept; the other
corrections are smaller than (or of the same order as) terms already
neglected. It can be checked that in the limit
$\varepsilon_{\lambda}\ll 1$ and $\varepsilon_{\rho}\ll 1$, Eqs.\
(\ref{eq3a.1}) and\ (\ref{eq3b.1}) coincide at first order in
$\varepsilon_{\rho}$ and third order in $\varepsilon_{\lambda}$, as it
should be.
%
%
%
\subsection{Uniform expressions in the regime:
$\varepsilon_{\lambda}\varepsilon_{\rho}\ll 1$}\label{sec3c}
The two previous regimes cover the whole domain
$\varepsilon_{\lambda}\varepsilon_{\rho}\equiv
\kappa_{(5)}^2r_b\langle\rho_m\rangle /6\ll 1$ (with, of course,
compatibility in the overlapping domain $\varepsilon_{\lambda}\ll 1$
and $\varepsilon_{\rho}\ll 1$ as mentioned at the end of the previous
section. This suggests the following uniform expressions valid in the
whole domain $\varepsilon_{\lambda}\varepsilon_{\rho}\ll 1$:
\begin{mathletters}\label{eq3c.1}
\begin{eqnarray}
&&H_{\langle a\rangle}^2=\frac{\kappa_{(5)}^2}{3}\,
\frac{\lbrack\alpha^2(\ell_{\lambda}/\ell_{\Lambda})+\eta\langle\rho_m\rangle
/(2\lambda)\rbrack}
{r_b\lbrack\alpha^2(\ell_{\lambda}/\ell_{\Lambda})+\eta\langle\rho_m\rangle
/\lambda\rbrack +\alpha^2\ell_{\lambda}}\,
\eta\langle\rho_m\rangle ,\label{eq3c.1a} \\
&&\dot{\langle\rho_m\rangle}+
3H_{\langle a\rangle}(\langle\rho_m\rangle+\langle p_m\rangle)=0,
\label{eq3c.1b}
\end{eqnarray}
\end{mathletters}
where
\begin{mathletters}\label{eq3c.2}
\begin{eqnarray}
&&\alpha^2 =\frac{\cos(k/2)}{\langle\cos^{1/2}(ky)\rangle^2}, \label{eq3c.2a} \\
&&\eta =\frac{\langle f(y)\cos(ky)\rangle}
{\langle f(y)\rangle\langle\cos^{1/2}(ky)\rangle^2}, \label{eq3c.2b}
\end{eqnarray}
\end{mathletters}
and
\begin{mathletters}\label{eq3c.3}
\begin{eqnarray}
&&k=2\left( 1+\frac{\langle\rho_m\rangle /\lambda}{1-u}\right)^{1/2}
\tan^{-1}\sqrt{\frac{u}{1-u}}, \label{eq3c.3a} \\
&&\frac{\ell_{\lambda}}{\ell_{\Lambda}}=
\frac{\sqrt{u(1-u)}}{\tan^{-1}\sqrt{u/(1-u)}}, \label{eq3c.3b} \\
&&\varepsilon_{\lambda}=\frac{\lbrack\tan^{-1}\sqrt{u/(1-u)}\rbrack^2}
{1-u}, \label{eq3c.3c}
\end{eqnarray}
\end{mathletters}
with $0\le u<1$.
%
%
\section{Effective cosmological equations at higher orders in the case of a
Minkowski bulk}\label{sec4}
In this section, we wish to extend the analysis of the previous section
to higher orders. In order to simplify the calculations, we will
restrict ourselves to the case where the bulk cosmological constant
vanishes ($\Lambda =0$), i.e. the spacetime outside the brane
effectively is a five-dimensional Minkowski spacetime. In this case,
one has $\ell_{\Lambda}^{-2}=0$ and Eq.\ (\ref{eq2.17}) reads
\begin{equation}\label{eq4.1}
H_{\langle a\rangle}^2=\frac{2(\alpha^2+\varepsilon\eta)}{r_b^2}
\left\lbrack
1-\sqrt{1-\frac{\varepsilon^2\eta^2+{\cal C}r_b^2/\langle a\rangle^4}
{(\alpha^2+\varepsilon\eta)^2}}\right\rbrack .
\end{equation}
As previously mentioned, in the limit of a vanishing brane thickness,
Eq.\ (\ref{eq4.1}) reduces to the unconventional Friedmann equation of
thin brane cosmology
\begin{equation}\label{eq4.2}
H_{\langle a\rangle}^2=\frac{\kappa_{(5)}^4}{36}\langle\rho_b\rangle^2
+\frac{{\cal C}}{\langle a\rangle^4},
\end{equation}
(N.B. $\alpha\rightarrow 1$ and $\eta\rightarrow 1$ as $r_b\rightarrow
0$). In the following, we will consider the limit $\varepsilon\ll 1$,
$r_b{\cal C}^{1/2}/\langle a\rangle^2=O(\varepsilon)$, and determine
the corrections to Eq.\ (\ref{eq4.2}) due to a finite brane thickness
(i.e. a finite $\varepsilon$).
%
%
\subsection{$r_b^2H_{\langle a\rangle}^2$ at third order in
$\varepsilon$}\label{sec4a}
At third order in $\varepsilon$, Eq.\ (\ref{eq4.1}) can be written as
\begin{equation}\label{eq4a.1}
r_b^2H_{\langle a\rangle}^2=
\frac{\varepsilon^2\eta^2+{\cal C}r_b^2/\langle a\rangle^4}
{\alpha^2+\varepsilon\eta} .
\end{equation}
At this order, $\alpha$ and $\eta$ must be expressed at first and zeroth
order respectively [note that $\eta$ behaves as $1+O(\varepsilon^2)$].
To determine $a$ perturbatively in $\varepsilon$ we will use the
$y$-expansion\ (\ref{eq3b.2}) with $\rho_b$ replacing $\rho_m$.
Inserting this expansion in Eq.\ (\ref{eq2.7}) with $\Lambda =0$, one
obtains
\begin{equation}\label{eq4a.2}
(\Sigma_a^2)''=-4\varepsilon\Sigma_{\rho}\Sigma_a^2+2r_b^2H_{\langle
a\rangle}^2.
\end{equation}
Since one needs $a$ at first order in $\varepsilon$, one can neglect the
second term on the right-hand side of Eq.\ (\ref{eq4a.2}) [which is
$O(\varepsilon^2 )$], and consider the constant (in $y$) component of
Eq.\ (\ref{eq4a.2}) only. One finds
$\overline{a}_1=-\varepsilon\overline{\rho}_0\overline{a}_0$ and
$a=\langle a\rangle\overline{a}_0(1-\varepsilon\overline{\rho}_0y^2)$.
In this expression, $\overline{\rho}_0$ must be determined at zeroth
order in $\varepsilon$, which yields $\overline{\rho}_0=1$ and
$\langle\rho_b\rangle$ solution to the usual 4D energy-momentum
conservation equation. Thus, at first order in $\varepsilon$ one has
$a=\langle a\rangle\lbrack 1-\varepsilon(y^2-1/12)\rbrack$, $\alpha
=1-\varepsilon /6$, and$\eta =1$. It follows that, at third order in
$\varepsilon$, the quantity $r_b^2H_{\langle a\rangle}^2$ reads
$$
r_b^2H_{\langle a\rangle}^2=
\frac{\varepsilon^2+{\cal C}r_b^2/\langle a\rangle^4}
{1+2\varepsilon/3},
$$
or, equivalently,
\begin{equation}\label{eq4a.3}
H_{\langle a\rangle}^2=
\frac{\kappa_{(5)}^4\langle\rho_b\rangle^2 /36
+{\cal C}/\langle a\rangle^4}
{1+\kappa_{(5)}^2r_b\langle\rho_b\rangle /9},
\end{equation}
with
\begin{equation}\label{eq4a.4}
\dot{\langle\rho_b\rangle}+
3H_{\langle a\rangle}
(\langle\rho_b\rangle +\langle p_b\rangle)=0 .
\end{equation}
%
%
\subsection{$r_b^2H_{\langle a\rangle}^2$ at fourth order in
$\varepsilon$}\label{sec4b}
At fourth order in $\varepsilon$, Eq.\ (\ref{eq4.1}) reads
\begin{eqnarray}
r_b^2H_{\langle a\rangle}^2&=&
\frac{\varepsilon^2\eta^2+{\cal C}r_b^2/\langle a\rangle^4}
{\alpha^2+\varepsilon\eta}
\left\lbrack 1+
\frac{\varepsilon^2\eta^2+{\cal C}r_b^2/\langle a\rangle^4}
{4(\alpha^2+\varepsilon\eta)^2}\right\rbrack \nonumber\\
&=&\frac{\varepsilon^2\eta^2+{\cal C}r_b^2/\langle a\rangle^4}
{\alpha^2+\varepsilon\eta}
\left\lbrack 1+\frac{1}{4}r_b^2H_{\langle a\rangle}^2+O(\varepsilon^3)
\right\rbrack ,\label{eq4b.1}
\end{eqnarray}
which gives
\begin{equation}\label{eq4b.2}
r_b^2H_{\langle a\rangle}^2=
\frac{\varepsilon^2\eta^2+{\cal C}r_b^2/\langle a\rangle^4}
{\alpha^2+\varepsilon\eta -
(\varepsilon^2\eta^2+{\cal C}r_b^2/\langle a\rangle^4)/4}.
\end{equation}
In the numerator $\eta$ must be determined at second order in
$\varepsilon$. In the denominator $\alpha$ must be determined at second
order in $\varepsilon$ and $\eta =1$ [N.B. $\eta =1+O(\varepsilon^2)$].
Since one needs $a$ at second order in $\varepsilon$, one must consider
the components of Eq.\ (\ref{eq4a.2}) up to terms proportional to
$y^2$. One finds
$\overline{a}_1=-\varepsilon\overline{\rho}_0\overline{a}_0
+r_b^2H_{\langle a\rangle}^2/(2\overline{a}_0)$ and
$\overline{a}_2=-(\varepsilon
/6)(2\overline{\rho}_0\overline{a}_1+\overline{\rho}_1\overline{a}_0)
-\overline{a}_1^2/(2\overline{a}_0)$, which gives after some
straightforward algebra
\begin{eqnarray*}
\frac{a}{\langle a\rangle}&=&1-\varepsilon
\left(y^2-\frac{1}{12}\right)+
\frac{{\cal C}r_b^2}{2\langle a\rangle^4}
\left(y^2-\frac{1}{12}\right) \\
&+&\frac{\varepsilon^2}{12}\left\lbrack 5\left(y^2-\frac{1}{12}\right)
-2\left(y^4-\frac{1}{80}\right)\right\rbrack \\
&+&\frac{\varepsilon\overline{\rho}_1}{12}\left\lbrack
\left(y^2-\frac{1}{12}\right) -2\left(y^4-\frac{1}{80}\right)
\right\rbrack .
\end{eqnarray*}
Thus, at second order in $\varepsilon$ one has
\begin{eqnarray*}
&&\alpha =1-\frac{\varepsilon}{6}+
\frac{{\cal C}r_b^2}{12\langle a\rangle^4}+
\frac{\varepsilon (11\varepsilon +\overline{\rho}_1)}{180}, \\
&&\eta =1+\frac{\varepsilon (\varepsilon -2\overline{\rho}_1)}
{180}.
\end{eqnarray*}
To obtain $\eta$ one has used the expression $\rho_b/\langle\rho_b\rangle
=1+\overline{\rho}_1(y^2-1/12)+O(\varepsilon^2)$ and the fact that the
$O(\varepsilon^2)$ term, the average of which must vanish, does not
contribute to $\langle\rho_b a^2\rangle$ at second order in
$\varepsilon$. Inserting these expressions of $\alpha$ and $\eta$ in\
(\ref{eq4b.2}), one obtains at fourth order in $\varepsilon$
\begin{equation}\label{eq4b.3}
r_b^2H_{\langle a\rangle}^2=
\frac{\varepsilon^2\lbrack 1+\varepsilon (\varepsilon
-2\overline{\rho}_1)/90\rbrack +{\cal C}r_b^2/\langle a\rangle^4}
{1+2\varepsilon /3-\varepsilon (9\varepsilon -\overline{\rho}_1)/90
-{\cal C}r_b^2/(12\langle a\rangle^4)}.
\end{equation}
The equation for $\langle\rho_b\rangle$ reads
\begin{equation}\label{eq4b.4}
\dot{\langle\rho_b\rangle}+3H_{\langle a\rangle}
(\langle\rho_b\rangle +\langle p_b\rangle )=
-3\left\langle
(\langle\rho_b\rangle\Sigma_{\rho}+
\langle p_b\rangle\Sigma_p)
\frac{\partial\log\Sigma_a}{\partial\tau}\right\rangle .
\end{equation}
Writing the right-hand side of this equation at second order in
$\varepsilon$, one obtains
\begin{equation}\label{eq4b.5}
\dot{\langle\rho_b\rangle}=
-\frac{3H_{\langle a\rangle}
(\langle\rho_b\rangle +\langle p_b\rangle )}
{1-\lbrack\varepsilon (\varepsilon +\overline{\rho}_1)+
\langle p_b\rangle\varepsilon (\varepsilon +\overline{p}_1)/
\langle\rho_b\rangle\rbrack /60}.
\end{equation}
To obtain this equation we have used $\Sigma_{\rho}=1+
\overline{\rho}_1(y^2-1/12)+O(\varepsilon^2)$, $\Sigma_p=1+
\overline{p}_1(y^2-1/12)+O(\varepsilon^2)$, and the fact that, at
second order in $\varepsilon$, there is no contribution of the
$O(\varepsilon^2)$ terms. It remains to determine the evolution
equation for $\overline{\rho}_1$ at lowest order in $\varepsilon$, i.e.
in which one keeps terms at most of first order. To this end let us
assume an equation of state of the form $p =w(\tau)\rho$, (which
implies $\Sigma_p=\Sigma_{\rho}$). In this case, the energy-momentum
conservation reads
\begin{equation}\label{eq4b.6}
\langle\rho_b\rangle\dot{\Sigma}_{\rho} +
\lbrack\dot{\langle\rho_b\rangle}+3H_{\langle a\rangle}
(1+w)\langle\rho_b\rangle\rbrack\Sigma_{\rho}=
-3(1+w)\langle\rho_b\rangle\Sigma_{\rho}
\frac{\partial\log\Sigma_a}{\partial\tau},
\end{equation}
which yields, at lowest order in $\varepsilon$,
\begin{equation}\label{eq4b.7}
\dot{\overline{\rho}}_1=-3(1+w)\dot{\overline{a}}_1
=3(1+w)\dot{\varepsilon}.
\end{equation}
Note that the second term on the left-hand side of Eq.\
(\ref{eq4b.6}) is second order and must be neglected, as it can
be seen from Eq.\ (\ref{eq4b.5}). If $w$ is constant over some time
interval $\tau_1\le\tau\le\tau_2$, one can integrate Eq.\
(\ref{eq4b.7}), which gives
\begin{equation}\label{rho12}
\overline{\rho}_1(\tau_1\le\tau\le\tau_2)=
\overline{\rho}_1(\tau_1)+3(1+w)
\lbrack\varepsilon (\tau)-\varepsilon (\tau_1)\rbrack .
\end{equation}
To summarize, at fourth order in $\varepsilon$, the quantity
$r_b^2H_{\langle a\rangle}^2$ reads [Eq.\ (\ref{eq4b.3})]
$$
r_b^2H_{\langle a\rangle}^2=
\frac{\varepsilon^2\lbrack 1+\varepsilon (\varepsilon
-2\overline{\rho}_1)/90\rbrack +{\cal C}r_b^2/\langle a\rangle^4}
{1+2\varepsilon /3-\varepsilon (9\varepsilon -\overline{\rho}_1)/90
-{\cal C}r_b^2/(12\langle a\rangle^4)},
$$
with [Eq.\ (\ref{eq4b.5})]
$$
\dot{\varepsilon}=-\frac{3H_{\langle a\rangle}(1+w)\varepsilon}
{1-\varepsilon\lbrack\varepsilon(1+w)+\overline{\rho}_1(1+w^2)\rbrack/60},
$$
and [Eqs.\ (\ref{eq4b.7}) and\ (\ref{eq4b.5})]
$$
\dot{\overline{\rho}}_1=9H_{\langle a\rangle}(1+w)^2\varepsilon .
$$
This shows that at this order, and for the next orders as well, one must
introduce auxiliary quantities, here ${\overline{\rho}}_1$, in order to
describe adequately the effective cosmology from the brane point of
view. This is a sign that one should not expect any effective
four-dimensional reduction to be possible in the regimes of order one
$\varepsilon$.
%
%
\section{Conclusion}\label{sec5}
In the present work, we have attempted to address in a phenomenological
way the question of finite brane thickness and its influence on the
effective four-dimensional cosmological equations. In order to simplify
our analysis, we have made a number of assumptions, which it might be
interesting to explore further or to relax in future investigations.
\paragraph*{}First we have defined in a heuristic way the notion of
four-dimensional quantities by simply integrating  over the fifth
coordinate $y$, which is defined in a Gaussian normal coordinate system
centered on the middle layer of the brane, supposed to be mirror
symmetric. Although this averaging procedure corresponds to the
standard one, one could maybe envisage, when having a warping factor,
other types of averaging.
\paragraph*{}Another important assumption was to take the brane
thickness constant in time. It would be worthwhile to generalize our
results when the brane thickness is allowed to evolve in time.
\paragraph*{}After emphasizing a few  limits of the present work, let us
now summarize the main results. A very instructive result of our work
is a generalized brane Friedmann equation, which interpolates between
the familiar Kaluza-Klein picture, where the effective Friedmann
equations after the averaging procedure yield exactly the usual
four-dimensional Friedmann equations, and the initially surprising thin
brane Friedmann equations, where the energy density enters
quadratically. We hope that our generalized Friedmann equation will
help in the intuitive understanding of the unconventional thin brane
result, and clarify the delimitations of the various regimes where
different Friedmann equations have to be applied.
\paragraph*{}Assuming the existence of a strict (in both patial and
temporal senses) cosmological constant, we have also obtained a
generalized version of the Randall-Sundrum condition of cancellation
between the bulk cosmological constant and the brane tension term in
the Friedmann equation. In our case, this cancellation condition
depends explicitly on the brane thickness and yields back the familiar
condition in the thin brane limit. Moreover, when this cancellation
condition is assumed to hold and when the cosmological matter content
of the brane is assumed to be small with respect to its tension, we
were able to solve for the brane profile and obtain the effective
cosmological equations.
\paragraph*{}Finally, we show that this is only in this case, as well as
in the limit where the brane thickness is sufficiently small, that one
can  make sense of a four-dimensional effective cosmology. In the other
cases, one can try to extend the regime of validity  of the effective
four-dimensional description but at the price of introducing auxiliary
quantities, which is the sign that the full five-dimensional
description is more adequate.
%
%
\acknowledgements
We wish to thank P. Bin\'etruy and N. Deruelle for very interesting
discussions.
%
%
\appendix
\section{Five-dimensional Einstein equations}\label{app}
\newcommand{\A}{A}
\newcommand{\B}{B}
\newcommand{\mmu}{\mu}
\newcommand{\mnu}{\nu}
\newcommand{\ii}{i}
\newcommand{\jj}{j}
\newcommand{\jl}{[}
\newcommand{\jr}{]}
\newcommand{\ml}{\sharp}
\newcommand{\mr}{\sharp}
\newcommand{\da}{\dot{a}}
\newcommand{\db}{\dot{b}}
\newcommand{\dn}{\dot{n}}
\newcommand{\dda}{\ddot{a}}
\newcommand{\ddb}{\ddot{b}}
\newcommand{\ddn}{\ddot{n}}
\newcommand{\pa}{a^{\prime}}
\newcommand{\pb}{b^{\prime}}
\newcommand{\pn}{n^{\prime}}
\newcommand{\ppa}{a^{\prime \prime}}
\newcommand{\ppb}{b^{\prime \prime}}
\newcommand{\ppn}{n^{\prime \prime}}
\newcommand{\fda}{\frac{\da}{a}}
\newcommand{\fdb}{\frac{\db}{b}}
\newcommand{\fdn}{\frac{\dn}{n}}
\newcommand{\fdda}{\frac{\dda}{a}}
\newcommand{\fddb}{\frac{\ddb}{b}}
\newcommand{\fddn}{\frac{\ddn}{n}}
\newcommand{\fpa}{\frac{\pa}{a}}
\newcommand{\fpb}{\frac{\pb}{b}}
\newcommand{\fpn}{\frac{\pn}{n}}
\newcommand{\fppa}{\frac{\ppa}{a}}
\newcommand{\fppb}{\frac{\ppb}{b}}
\newcommand{\fppn}{\frac{\ppn}{n}}
Inserting the ansatz (\ref{eq2.1}) for the metric, the non-vanishing
components of the Einstein tensor $G_{AB}$ corresponding to the antsatz
(\ref{eq2.1}) for the metric are:
\begin{eqnarray}
{ G}_{00} &=& 3\left\{ \fda \left( \fda+ \fdb \right) - \frac{n^2}{r_b^2}
\left(\fppa + \fpa \left( \fpa - \fpb \right) \right)  \right\},
\label{ein00} \\
 { G}_{\ii\jj} &=&
\frac{a^2}{b^2} \delta_{ij}\left\{\fpa
\left(\fpa+2\fpn\right)
+2\fppa+\fppn\right\}
\nonumber \\
& &+\frac{a^2}{n^2} \delta_{ij} \left\{ \fda \left(-\fda+2\fdn\right)-2\fdda
 \right\} ,
\label{einij} \\
{ G}_{05} &=&  3\left(\fpn \fda  - \frac{\dot{a}^{\prime}}{a}
 \right),
\label{ein05} \\
{ G}_{55} &=& 3\left\{ \fpa \left(\fpa+\fpn \right) - \frac{r_b^2}{n^2}
\left(\fda \left(\fda-\fdn \right) + \fdda\right) \right\}.
\label{ein55}
\end{eqnarray}
In the above expressions, a prime stands for a derivative with respect
to $y$, and a dot for a derivative with respect to $\tau$.
%
%

%
%
\begin{figure}\caption{Parameter $k$ (solid line) and $\varepsilon_{\lambda}$
(dashed-dotted line) as  functions of $u$.}\label{fig1}\end{figure}
\begin{figure}\caption{Variation with $k/k_{max}$ of the parameters
${\cal A}$, $\tilde{\eta}$ and  $\alpha$ (from top to
bottom).}\label{fig2}\end{figure}
\newpage\centerline{\epsfxsize=4in \epsffile{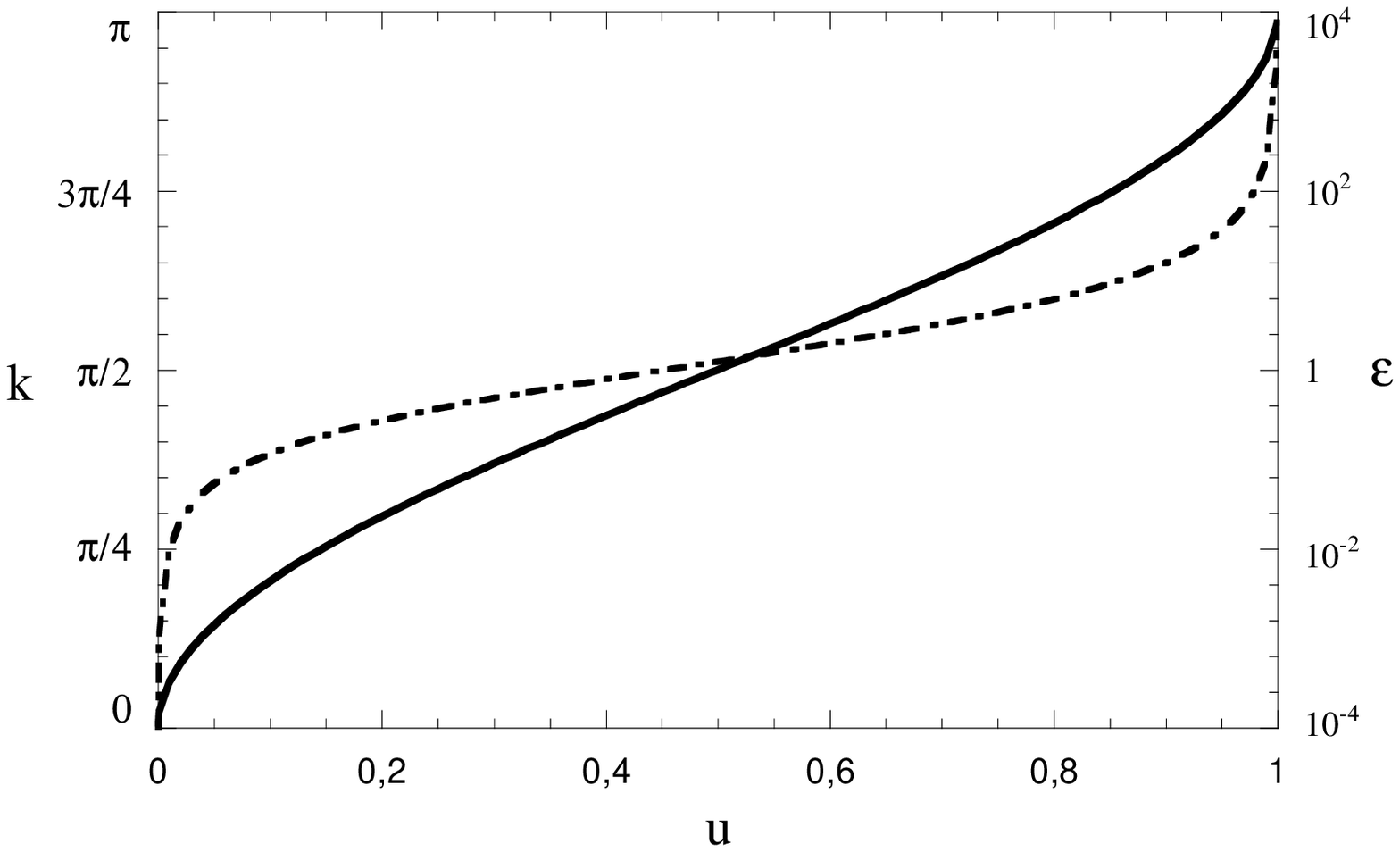}}
\centerline{Figure \ref{fig1}.}
\newpage\centerline{\epsfxsize=4in \epsffile{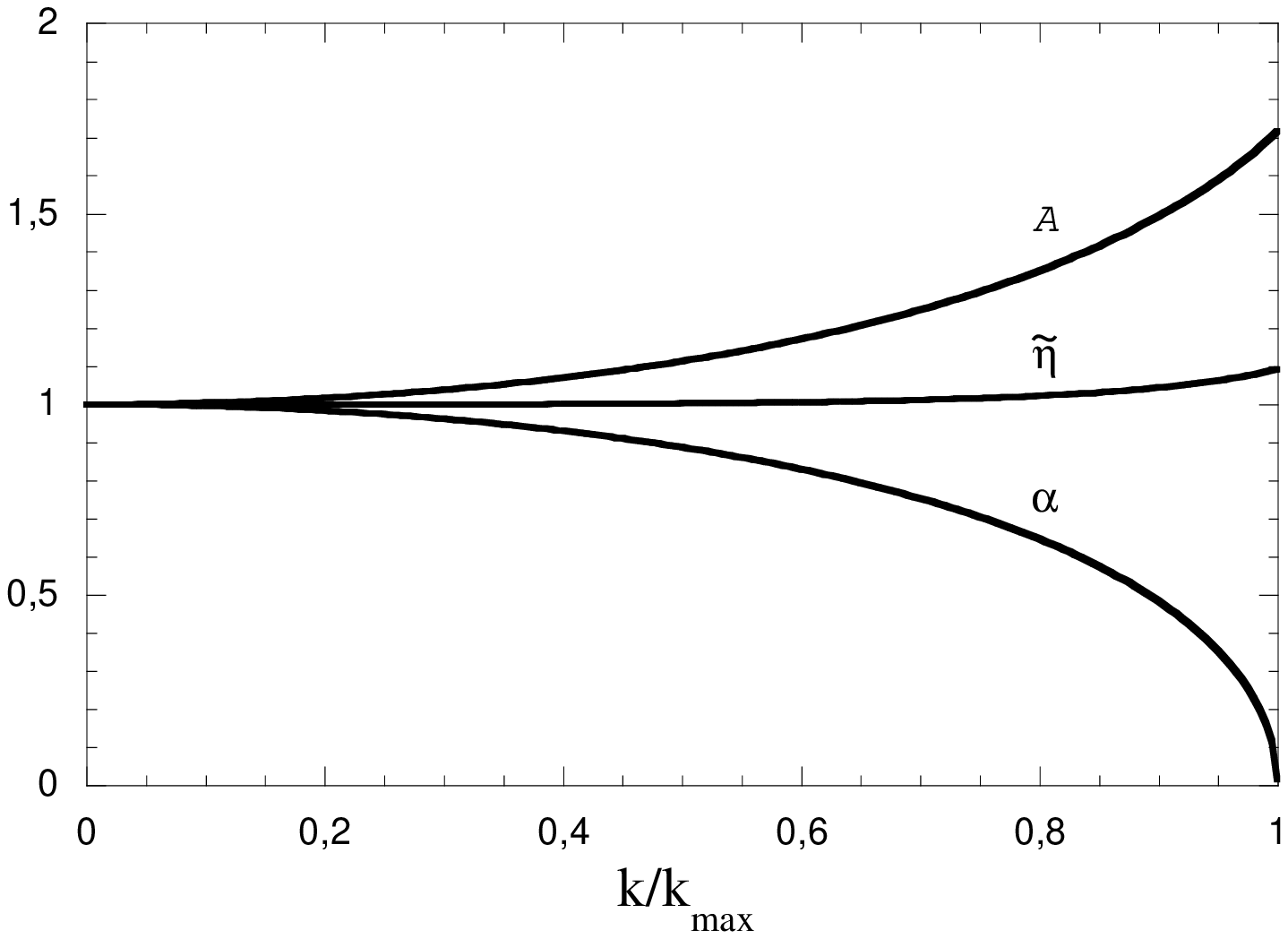}}
\centerline{Figure \ref{fig2}.}
\end{document}